\documentclass[runningheads]{llncs}

\usepackage[T1]{fontenc}
\def\doi#1{\href{https://doi.org/\detokenize{#1}}{\url{https://doi.org/\detokenize{#1}}}}
\usepackage{graphicx}
\usepackage{listings}
\usepackage{amsmath}
\usepackage{amssymb}
\usepackage{graphicx}
\usepackage{bm}
\usepackage{color}
\usepackage{authblk}
\usepackage{booktabs}
\usepackage{multirow}
\usepackage{float}
\usepackage{ctable}
\usepackage[colorlinks, citecolor=green]{hyperref}
\begin{document}

\title{Global k-Space Interpolation for Dynamic MRI Reconstruction using Masked Image Modeling}

 \author{Jiazhen Pan\inst{1}, Suprosanna Shit\inst{1}, Özgün Turgut\inst{1}, \authorcr Wenqi Huang\inst{1}, Hongwei~Bran~Li\inst{1,2}, Nil Stolt-Ansó\inst{3}, \authorcr Thomas Küstner\inst{4}, Kerstin~Hammernik\inst{3,5}, Daniel Rueckert\inst{1,3,5}}
 \authorrunning{Pan et al.}
 
 %
 \institute{
School of Medicine, Klinikum Rechts der Isar, Technical University of Munich, Germany
\and
Department of Quantitative Biomedicine, University of Zurich, Switzerland
\and
School of Computation, Information and Technology, Technical University of Munich, Germany
\and
Medical Image And Data Analysis, University Hospital of Tübingen, Germany
\and
Department of Computing, Imperial College London, United Kingdom
\\
\email{\{jiazhen.pan,suprosanna.shit\}@tum.de}
}
\titlerunning{Accepted at MICCAI 2023}

\maketitle              

\begin{abstract}

In dynamic Magnetic Resonance Imaging (MRI), k-space is typically undersampled due to limited scan time, resulting in aliasing artifacts in the image domain. Hence, dynamic MR reconstruction requires not only modeling spatial frequency components in the x and y directions of k-space but also considering temporal redundancy. Most previous works rely on image-domain regularizers (priors) to conduct MR reconstruction. In contrast, we focus on interpolating the undersampled k-space before obtaining images with Fourier transform. In this work, we connect masked image modeling with k-space interpolation and propose a novel Transformer-based k-space Global Interpolation Network, termed k-GIN. Our k-GIN learns global dependencies among low- and high-frequency components of 2D+t k-space and uses it to interpolate unsampled data. Further, we propose a novel k-space Iterative Refinement Module (k-IRM) to enhance the high-frequency components learning. We evaluate our approach on 92 in-house 2D+t cardiac MR subjects and compare it to MR reconstruction methods with image-domain regularizers. Experiments show that our proposed k-space interpolation method quantitatively and qualitatively outperforms baseline methods. Importantly, the proposed approach achieves substantially higher robustness and generalizability in cases of highly-undersampled MR data. For video presentation, poster, GIF results and code please check our project page: \url{https://jzpeterpan.github.io/k-gin.github.io/}. 


\end{abstract}

\section{Introduction}
CINE Cardiac Magnetic Resonance (CMR) imaging is widely recognized as the gold standard for evaluating cardiac morphology and function~\cite{Lee2018}. 
Raw data for CMR is acquired in the frequency domain (k-space). MR reconstruction from k-space data with high spatio-temporal resolutions throughout the cardiac cycle is an essential step for CMR. Short scan times, ideally within a single breath-hold, are preferable to minimize patient discomfort and reduce potential image artifacts caused by patient motion. Typically, due to the restricted scanning times, only a limited amount of k-space data can be obtained for each temporal frame. Note that while some k-space data are unsampled, the sampled ones are reliable sources of information. However, the Fourier transform of undersampled k-space corrupts a broad region of pixels in the image domain with aliasing artifacts because of violating the Nyquist–Shannon sampling theorem. Previous works~\cite{LplusS,ktslr,DeepInverse,VN-Net,Schlemper2018} have attempted to remove the image artifacts primarily by regularizing on the image domain using conventional/learning-based image priors. However, after the Fourier transform the artifacts may completely distort and/or obscure tissues of interest before the image-domain regularizers kick in, making these methods challenging to recover true tissue structures.

On a different principle, k-space interpolation methods first attempt to estimate the full k-space leveraging redundancy in sampled frequency components before Fourier transform. Image domain methods rely on artifacts-specific image priors to denoise the corrupted pixels, making them susceptible to variability in artifact types arising from different undersampling factors. Unlike image domain methods, k-space-based methods have a consistent task of interpolating missing data from reliable sampled ones, even though the undersampling factor may vary. This makes k-space interpolation methods simple, robust and generic over multiple undersampling factors. 

In this work, we are interested in learning an entirely k-space-based interpolation for the Cartesian undersampled dynamic MR data. An accurate learnable k-space interpolator can be achieved via (1) \emph{a rich representation of the sampled k-space data}, which can facilitate the exploitation of the limited available samples, and (2) \emph{global dependency modeling of k-space} to interpolate unsampled data from the learned representation.  Modeling global dependencies are beneficial because a local structure in the image domain is represented by a wide range of frequency components in k-space. Furthermore, in the context of dynamic MR, the interpolator also has to exploit temporal redundancies. 

In the recent past, masked image modeling~\cite{simmim,MaskedAutoencoders2021} has emerged as a promising method for learning rich generalizable representation by reconstructing the whole image from a masked (undersampled) input. Masked Autoencoders (MAE)~\cite{MaskedAutoencoders2021} are one such model that leverages the \emph{global dependencies} of the undersampled input using Transformers and learns masked-based \emph{rich feature representation}. Despite sharing the same reconstruction principle, MAE has not been explored in k-space interpolation of Cartesian undersampled data. In this work, we cast 2D+t k-space interpolation as a masked signal reconstruction problem and propose a novel Transformer-based method entirely in k-space. Further, we introduce a refinement module on k-space to boost the accuracy of high-frequency interpolation. \textbf{Our contributions can be summarized as follows}: 
\begin{enumerate}
    \item We propose a novel k-space Global Interpolation Network, termed k-GIN, leveraging masked image modeling for the first time in k-space. To the best of our knowledge, our work enables the first Transformer-based k-space interpolation for 2D+t MR reconstruction.
    \item Next, we propose k-space Iterative Refinement Module, termed k-IRM that refines k-GIN interpolation by efficiently gathering spatio-temporal redundancy of the MR data. Crucially, k-IRM specializes in learning high-frequency details with the aid of customized High-Dynamic-Range (HDR) loss. 
    \item We evaluate our approach on 92 in-house CMR subjects and compare it to model-based reconstruction baselines using image priors. Our experiments show that the proposed k-space interpolator outperforms baseline methods with superior qualitative and quantitative results. Importantly, our method demonstrates improved robustness and generalizability regarding varying undersampling factors than the model-based counterparts.
\end{enumerate}

\section{Related work}

\paragraph{\textbf{Reconstruction with image priors}} is broadly used with either image-only denoising~\cite{residualDenoising,dagan,cnn_sr} or with model-based approaches to incorporate the k-space consistency by solving an inverse problem. For the latter, the physics-based model can be formulated as a low-rank and a sparse matrix decomposition in CMR~\cite{LplusS,Huang21}, or a motion-compensated MR reconstruction problem~\cite{Batchelor2005,Pan2022-1,Pan2023}, or data consistency terms with convolution-based~\cite{Schlemper2018,VN-Net} or Transformers-based~\cite{SwinMRI,Zeroshot} image regularizers.

\paragraph{\textbf{k-space-domain interpolation}} methods include works~\cite{GRAPPA,lustig2010spirit}, which have introduced auto-calibration signals (ACS) in the multi-coil k-space center of Cartesian sampled data. RAKI~\cite{RAKI,kim2019loraki} uses convolutional networks for optimizing imaging and scanner-specific protocols. Nevertheless, these methods have limited flexibility during the scanning process since they all require a fixed set of ACS data in k-space. \cite{sake,aloha} introduced k-space interpolation methods which do not require calibration signals. \cite{KspaceRecon} proposed a k-space U-Net under residual learning setting. However, these methods heavily rely on local operators such as convolution and may overlook non-local redundancies. \cite{ProjectionTransformer} uses Transformers applicable only on radial sampled k-space data. However, using masked image modeling with Transformers in k-space for dynamic MR imaging e.g. 2D+t CMR data has not been studied yet.

\paragraph{\textbf{Hybrid approaches}} combine information from both k-space and image-domain. 
KIKI-Net~\cite{KIKINET} employs an alternating optimization between the image domain and k-space. \cite{CDFNET,JointKandI} use parallel architectures for k-space and image-domain simultaneously. However, their ablation shows limited contribution coming from the k-space compared to the image domain, implying an under-exploitation of the k-space. Concurrently, \cite{zhao2022kspaceTransformer} use Transformers in k-space but their performance is heavily dependent on the image domain fine-tuning at the final stage.

\section{Method}
\label{sec:meth}
\begin{figure}[!t]
    \centering
    \includegraphics[width=\linewidth]{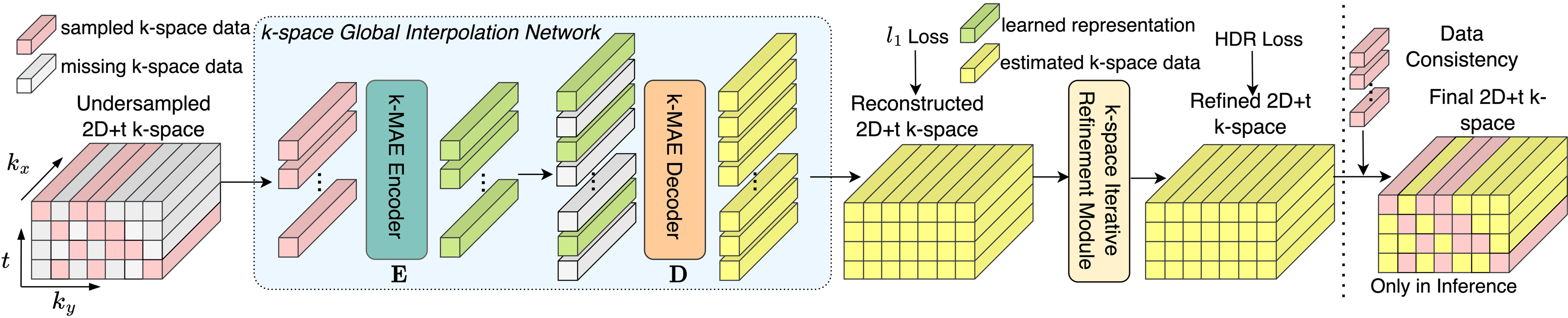}
    \caption{The proposed k-space-based dynamic MR reconstruction framework consists of k-space Global Interpolation Network (k-GIN) (see \ref{ssec:k_mae}) and k-space Iterative Refinement Module (k-IRM) for refining the k-GIN interpolation (see \ref{ssec:k_irm}). In the final stage for the inference, we replace k-space estimation at the sampled position with ground-truth k-space values, ensuring the data consistency.}
    \label{fig:k_mae}
\end{figure}
Fully sampled complex 2D+t dynamic MR k-space data can be expressed as $\mathbf{y} \in \mathbb{C}^{XYT}$, $X$ and $Y$ are the height ($k_x$) and the width ($k_y$) of the k-space matrix, and $T$ is the number of frames along time. In this work, we express k-space as 2 channels (real and imaginary) data $\mathbf{y} \in \mathbb{R}^{2XYT}$. For the MR acquisition, a binary Cartesian sampling mask $M \in \mathbb{Z}^{YT} |M_{ij}\in \{0,1\}$ is applied in the $k_{y}\text{-}t$ plane, i.e. all k-space values along the $k_{x}$ (readout direction) are sampled if the mask is 1, and remains unsampled if the mask is 0. Fig.~\ref{fig:k_mae} shows a pictorial undersampled k-space data. Let us denote the collection of sampled k-space lines as $\mathbf{y}_s$ and unsampled lines as $\mathbf{y}_u$. The dynamic MR reconstruction task is to estimate $\mathbf{y}_u$ and reconstruct $\mathbf{y}$ using $\mathbf{y}_s$ only. In this work, we propose a novel Transformer-based reconstruction framework consisting of 1) k-GIN to learn global representation and 2) k-IRM to achieve refined k-space interpolation with a focus on high-frequency components. 

\subsection{k-space Global Interpolation Network (k-GIN)}
\label{ssec:k_mae}
In our proposed approach, we work on the $k_{y}\text{-}t$ plane and consider $k_x$ as the channel dimension. Further, we propose each point in the $k_{y}\text{-}t$ plane to be an individual \emph{token}. In total, we have $YT$ number of tokens, out of which $YT/R$ are sampled tokens for an undersampling factor of $R$. 
Our objective is to contextualize global dependencies among every sampled token. For that, we use a ViT/MAE~\cite{ViT,MaskedAutoencoders2021} encoder $\mathbf{E}$ consisting of alternating blocks of multi-head self-attention and multi-layer-perceptrons. The encoder takes advantage of each token's position embedding to correctly attribute its location. Following ViT, we use LayerNorm and GELU activation. We obtain rich feature representation $\mathbf{f}_E = \mathbf{E}(\mathbf{y}_s)$ of the sampled k-space from the encoder.

Next, we want a preliminary estimate of the undersampled k-space data from the learned feature representation. To this end, we employ a decoder $\mathbf{D}$ of similar architecture as the encoder. We initialize all the unsampled tokens $\mathbf{y}_u$ with a single learnable token shared among them. Subsequently, we add their corresponding position embedding to these unsampled tokens. During the decoding process, the unsampled tokens attend to the well-contextualized features $\mathbf{f}_E$ and produce an estimate of the whole k-space $\mathbf{\hat{y}}_r = \mathbf{D}\left([\mathbf{f}_E, \mathbf{y}_u]\right)$.  Since our masking pattern includes more sampled data in low-frequency than high-frequency components, we observe k-GIN gradually learn from low-frequency to high-frequency. Note that the imbalance of magnitude in k-space results in more emphasis on low-frequency when $\ell_1$ loss is applied between estimation and ground-truth. We leverage this property into the learning behavior of k-GIN and deliberately use $\ell_1$ loss between $\mathbf{\hat{y}}_r$ and $\mathbf{y}$, read as $\mathcal{L}_{\ell_1} = \left\|\hat{\mathbf{y}}_r - \mathbf{y}\right\|_1$.

\begin{figure}[!t]
    \centering
    \includegraphics[width=0.9\linewidth]{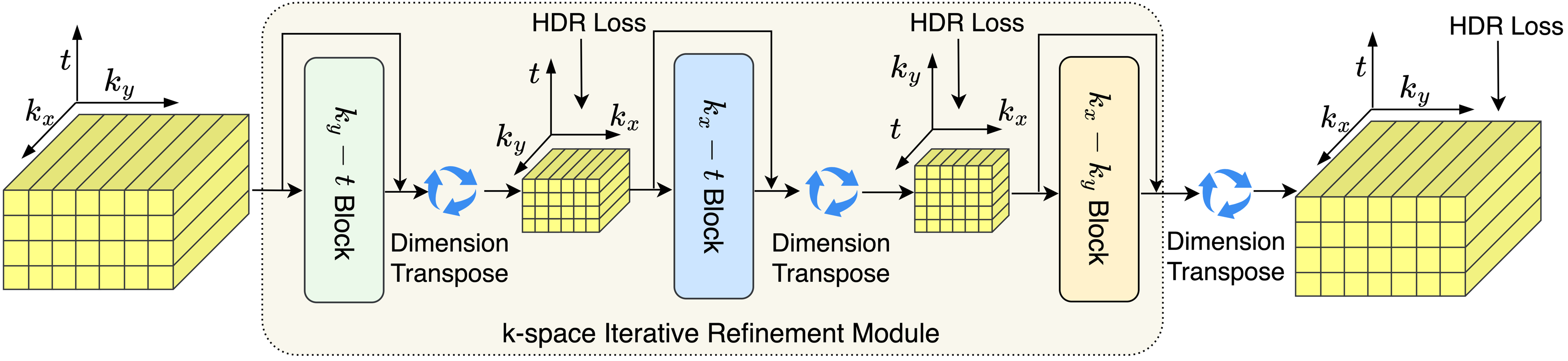}
    \caption{k-space Iterative Refinement Module (k-IRM) refines high-frequency components of the k-space data (see \ref{ssec:k_mae}). Its refinement Transformer blocks extract the spatio-temporal redundancy by operating the on $k_y\text{-}t$, $k_x\text{-}t$ and $k_x\text{-}k_y$ plane.}
    \label{fig:k_irm}
\end{figure}

\subsection{k-space Iterative Refinement Module (k-IRM)}
\label{ssec:k_irm}
Using $\ell_1$ loss in k-GIN makes it focus more on the low-frequency components learning but the high-frequency estimation is still sub-optimal. Inspired by the iterative refinement strategy~\cite{dedetr} which is widely used to improve estimation performance, we propose to augment k-GIN's expressive power, especially in high-frequency components, with k-space Iterative Refinement Module. This consists of three Transformer blocks that operate on three orthogonal planes. All three blocks are identical in architecture. The first block operates on $k_{y}\text{-}t$ plane and treats $k_x$ as channel dimension. The second block operates on $k_{x}\text{-}t$ plane and considers $k_y$ as channels, while the final block operates on $k_x\text{-}k_y$ plane with $t$ as channel dimension. Note that for the final refinement block uses $4 \times 4$ size token while the previous two blocks consider each point as a single token. These configurations enable scalable exploration of the spatio-temporal redundancy present in the output of the k-GIN. We denote $\hat{\mathbf{y}}_1$, $\hat{\mathbf{y}}_2$ and $\hat{\mathbf{y}}_3$ as the estimation after each block in the k-IRM. We apply the skip connection between each refinement block and iteratively minimize the residual error at each stage.

Inspired by~\cite{DarkNerf,huang2023neural}, we applied an approximated logarithm loss function called High-Dynamic Range (HDR) loss for all the stages of k-IRM. HDR loss handles the large magnitude difference in the k-space data and makes the network pay more attention to high-frequency learning. The HDR loss function is defined as: 
$
    \mathcal{L}_{\mbox{HDR}}= \sum_{i=1}^{3}\left\|\frac{\hat{\mathbf{y}}_i - \mathbf{y}}{s(\hat{\mathbf{y}})+\epsilon}\right\|^2_2
    $
where $s(\cdot)$ is the stop-gradient operator preventing the network back-propagation of estimation in the denominator and $\epsilon$ controls the operational range of the logarithmic approximation.

\subsection{Inference}
The inference is identical to the training till obtaining the refined output from the k-IRM. Then we replace k-space estimation at the sampled position with ground-truth k-space values, ensuring the data-consistency. Note that this step is not done during the training as it deteriorates learning k-space representation. Once full k-space has been estimated, we use Fourier transform to obtain the image reconstruction during the inference. Note that image reconstruction is not needed during the training since our framework is entirely based in k-space.

\section{Data and Experiments}

\paragraph{\textbf{Dataset.}}
The training was performed on 81 subjects (a mix of patients and healthy subjects) of in-house acquired short-axis 2D CINE CMR, whereas testing was carried out on 11 subjects. Data were acquired with 30/34 multiple receiver coils and 2D balanced steady-state free precession sequence on a 1.5T MR (Siemens Aera with TE=1.06 ms, TR=2.12 ms, resolution=$1.9{\times}1.9\text{mm}^2$ with 8mm slice thickness, 8 breath-holds of 15s duration). The MR data were acquired with a matrix size of $192{\times}156$ with 25 temporal cardiac phases (40ms temporal resolution). Afterwards, these data were converted to single-coil MR imaging and k-space data using coil sensitivity maps, simulating a fully sampled single-coil acquisition. A stack of 12 slices along the long axis was collected, resulting in 415/86 image sequence (2D+t) for training/test. 
\paragraph{\textbf{Implementation Details.}}
We use an NVIDIA A6000 GPU to train our framework. The batch size was set to 1 with a one-cycle learning-rate scheduler (max. learning rate 0.0001). We use 8 layers, 8 heads and 512 embedding dimensions for all of our Transformer blocks.  We train our network with joint $\ell_1$ and  HDR-loss with  $\epsilon$ tuned to 0.5. Training and inference were carried out on retrospectively undersampled images with masks randomly generated by VISTA~\cite{Vista}. We train the network with $R=4$ undersampled data while we test our method on an undersampled factor $R=$ 4, 6 and 8 during the inference. We can use this inference strategy to test our model's generalizability and robustness to different undersampling factors in comparison to the following baseline methods. 
\paragraph{\textbf{Baseline Methods and Metrics.}}
We compare the proposed framework with three single-coil MR reconstruction methods that apply image priors: TV-norm Optimization (TV-Optim) which is widely used in reconstruction~\cite{ktslr,tvmri}, L+S~\cite{LplusS} and DcCNN~\cite{Schlemper2018}. TV-Optim reconstructs the image using TV-norm~\cite{tv} as the image regularizer. L+S leverages compressed sensing techniques and addresses the reconstruction using low rankness and sparsity of the CMR as the image prior, whilst DcCNN employs 3D convolutional neural networks in the image domain together with data-consistency terms. We use the same training and inference strategy for DcCNN to test its model robustness.

We Fourier transform our interpolated full k-space to obtain the image reconstruction and utilize Peak Signal-to-Noise Ratio (PSNR), Structural Similarity Index (SSIM) and Normalized Mean Squared Error (NMSE) to evaluate the reconstruction performance with the baseline quantitatively.

\section{Results and Discussion}

\begin{table}[!t]
    \centering
    \setlength{\tabcolsep}{3mm}{}
    \caption{Quantitative analysis of reconstruction for accelerated CINE CMR (R=4, 6 and 8) using TV-Optim, L+S~\cite{LplusS}, DcCNN~\cite{Schlemper2018} and the proposed method. PSNR (sequence based), SSIM and NMSE are used to evaluate the reconstruction performance. The mean value with the standard deviations are shown. The best results are marked in bold.}
    \begin{tabular}{ccccc}
        \toprule
        Acc R & Methods & NMSE~$\downarrow$ & SSIM~$\uparrow$ & PSNR~$\uparrow$ \\ \midrule
        \multirow{4}*{4} 
        & TV-Optim & $0.120 \pm 0.031$ & $0.922 \pm 0.026$ & $36.743 \pm 3.233$ \\
        ~ & L+S~\cite{LplusS} & $0.097 \pm 0.026$ & $0.949 \pm 0.021$ & $39.346 \pm 2.911$\\
        ~ & DcCNN~\cite{Schlemper2018} & $\mathbf{0.087 \pm 0.022} $ & $0.957 \pm 0.019 $ & $40.293 \pm 2.891$\\
        ~ & Proposed & $0.088 \pm 0.022 $ & $\mathbf{0.958 \pm 0.019}$ & $\mathbf{40.368 \pm 3.030}$\\  \midrule
        \multirow{4}*{6} 
          & TV-Optim & $ 0.161 \pm 0.047 $ & $0.887 \pm 0.040$ & $34.066 \pm 3.605$  \\
        ~ & L+S & $0.154 \pm 0.042$ & $0.901 \pm 0.036 $ & $34.921 \pm 3.174$ \\
        ~ & DcCNN & $0.116 \pm 0.029$ &  $0.932 \pm 0.026$ & $37.666 \pm 2.768$ \\
        ~ & Proposed & $\mathbf{0.109 \pm 0.029}$ & $\mathbf{0.940 \pm 0.026}$ & $\mathbf{38.461 \pm 3.095}$ \\ \midrule
        \multirow{4}*{8} 
          & TV-Optim & $0.289 \pm 0.078$ & $0.808 \pm 0.067$  & $29.052 \pm 3.817$ \\
        ~ & L+S & $0.245 \pm 0.061$ & $0.826 \pm 0.047$ & $30.413 \pm 2.888$ \\
        ~ & DcCNN & $0.276 \pm 0.047$ & $0.821 \pm 0.040$ & $29.778 \pm 2.822$ \\
        ~ & Proposed & $\mathbf{0.151 \pm 0.041}$ & $\mathbf{0.904 \pm 0.036}$ & $\mathbf{35.674 \pm 3.293}$ \\ \bottomrule
    \end{tabular}
    \label{tab:main}
\end{table}

\begin{figure}[t!]
\includegraphics[width=\textwidth]{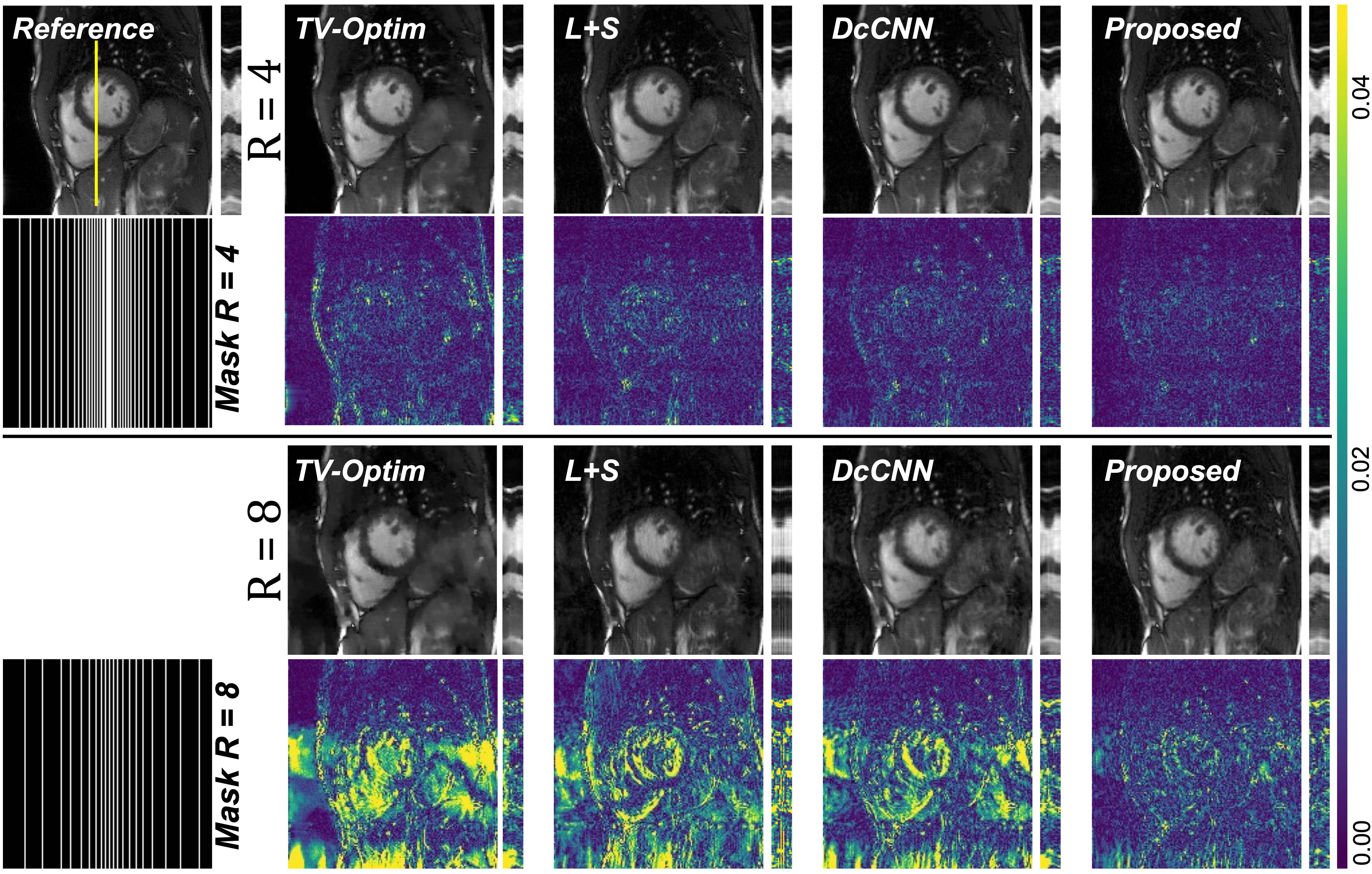}
\caption{Qualitative comparison of the proposed method with TV-Optim, L+S and DcCNN in the $R = 4~\text{and}~8$ undersampled data. Reference images, undersampling masks, reconstructed ($x\text{-}y$ and $y\text{-}t$ plane) images and their corresponding error maps are showcased. The selected y-axis is marked with a yellow line in the reference image.} \label{f2}
\end{figure}

The quantitative results in Table~\ref{tab:main} show consistent superior performance of the proposed method across every single undersampling factor compared to all other baseline methods. Fig.~\ref{f2} shows a qualitative comparison for a typical test sample. It can be seen that the reconstruction methods with image priors can still provide comparable results at $R=4$, however, suffer from a large performance drop when acceleration rates get higher, especially at $R=8$. The non-trivial hyper-parameters tuning has to be carried out for L+S and TV-Optim to adapt to the specific image prior at different acceleration factors. It is also noteworthy that the proposed method and DcCNN are both trained only on $R=4$ undersampled CMR. 
DcCNN demonstrates inferior reconstruction for $R=8$ since there is a mismatch in artifact characteristics between $R=4$ and $R=8$. On the contrary, the task of interpolating k-space for $R=4$ and $R=8$ remains the same, i.e., to estimate missing data from sampled data. We efficiently leverage rich contextualized representation of k-GIN to interpolate full k-space even when a lesser number of sampled k-space data are given as input than seen during training. The observation confirms the superior robustness and generalizability of our proposed framework.


Next, we conduct an ablation study to validate our architectural design. We carry out experiments to investigate the impact of applying k-IRM. We conduct the interpolation using 1) only k-GIN, 2) k-GIN + $k_y\text{-}t$ plane refinement, 3) k-GIN + $k_x\text{-}t$ plane refinement, 4) k-GIN + $k_x\text{-}k_y$ plane refinement and 5) k-GIN with all three refinement blocks. Table~2 in supplementary presents quantitative comparisons amongst five configurations as above. We observe k-IRM offers the best performance when all 3 refinement blocks are used together. In the second ablation, Table~3 in supplementary shows the usefulness of applying $\ell_1$ in k-GIN and HDR in k-IRM. HDR makes k-GIN's learning inefficient since HDR deviates from its learning principle of "first low-frequency then high-frequency". On the other hand, $\ell_1+\ell_1$ combination hinders high-frequency learning. 

\paragraph{\textbf{Outlook.}} Previous works~\cite{KIKINET,CDFNET} have speculated limited usefulness coming from k-space in a hybrid setting. However, our work presents strong evidence of k-space representation power which can be leveraged in future work with hybrid reconstruction setup. Furthermore, one can utilize our work as a pre-training task since the image reconstruction itself is an "intermediate step" for downstream tasks e.g. cardiac segmentation and disease classification. In the future, one can reuse the learned encoder representation of k-GIN to directly solve downstream tasks without requiring image reconstruction.

\paragraph{\textbf{Limitation.}} We also acknowledge some limitations of the work. First, we have not evaluated our method on prospectively collected data, which would be our focus in future work. Second, the current study only investigates the single coil setup due to hardware memory limitations. In the future, we will address the multi-coil scenario by applying more memory-efficient Transformers backbones e.g.~\cite{dao2022flashattention,xiong2021nystromformer}.

\section{Conclusion}
In this work, we proposed a novel Transformer-based method with mask image modeling to solve the dynamic CMR reconstruction by only interpolating the k-space without any image-domain priors. Our framework leverages Transformers' global dependencies to exploit redundancies in all three $k_x$-, $k_y$- and t-domain. Additionally, we proposed a novel refinement module (k-IRM) to boost high-frequency learning in k-space. Together, k-GIN and k-IRM not only produce high-quality k-space interpolation and superior CMR reconstruction but also generalize significantly better than baselines for higher undersampling factors.

\section{Acknowledgements}
This work is partly supported by the European Research Council (ERC) with Grant Agreement no. 884622. Suprosanna Shit is supported by ERC with the Horizon 2020 research and innovation program (101045128-iBack-epic-ERC2021-COG). Hongwei Bran Li is supported by an Nvidia GPU research grant.

\bibliographystyle{splncs04}

\bibliography{bibliography.bib}
\clearpage
\appendix
\end{document}